\documentclass[12pt]{article}
\usepackage{fancyhdr}

\usepackage{mathrsfs}
\usepackage[T1]{fontenc}
\usepackage{mathpazo}
\usepackage{setspace}
\usepackage{amsfonts}
\usepackage{amssymb}
\usepackage{amsmath}
\usepackage{epsfig}
\usepackage{latexsym}
\usepackage{color}
\usepackage{graphicx}
\usepackage{nicefrac}
\usepackage[latin1]{inputenc}
\usepackage{slashed}
\usepackage{multirow}
\usepackage{cite}
\usepackage{comment}
\usepackage{hyperref}
\def\hybrid{\topmargin -20pt    \oddsidemargin 0pt
        \headheight 0pt \headsep 0pt
        \textwidth 6.25in       % A4 paper
        \textheight 9.25in       % A4 paper
        \marginparwidth .875in
        \parskip 5pt plus 1pt   \jot = 1.5ex}

\hybrid

\def\baselinestretch{1.2}

\catcode`\@=11

\def\marginnote#1{}
\newcount\hour
\newcount\minute
\newtoks\amorpm
\hour=\time\divide\hour by60
\minute=\time{\multiply\hour by60 \global\advance\minute by-\hour}
\edef\standardtime{{\ifnum\hour<12 \global\amorpm={am}%
        \else\global\amorpm={pm}\advance\hour by-12 \fi
        \ifnum\hour=0 \hour=12 \fi
        \number\hour:\ifnum\minute<10 0\fi\number\minute\the\amorpm}}
\edef\militarytime{\number\hour:\ifnum\minute<10 0\fi\number\minute}
\def\draftlabel#1{{\@bsphack\if@filesw {\let\thepage\relax
   \xdef\@gtempa{\write\@auxout{\string
      \newlabel{#1}{{\@currentlabel}{\thepage}}}}}\@gtempa
   \if@nobreak \ifvmode\nobreak\fi\fi\fi\@esphack}
        \gdef\@eqnlabel{#1}}
\def\@eqnlabel{}
\def\@vacuum{}
\def\draftmarginnote#1{\marginpar{\raggedright\scriptsize\tt#1}}

\def\draft{\oddsidemargin -.5truein
        \def\@oddfoot{\sl preliminary draft \hfil
        \rm\thepage\hfil\sl\today\quad\militarytime}
        \let\@evenfoot\@oddfoot \overfullrule 3pt
        \let\label=\draftlabel
        \let\marginnote=\draftmarginnote
   \def\@eqnnum{(\theequation)\rlap{\kern\marginparsep\tt\@eqnlabel}%
\global\let\@eqnlabel\@vacuum}  }

\def\preprint{\twocolumn\sloppy\flushbottom\parindent 2em
        \leftmargini 2em\leftmarginv .5em\leftmarginvi .5em
        \oddsidemargin -.5in    \evensidemargin -.5in
        \columnsep .4in \footheight 0pt
        \textwidth 10.in        \topmargin  -.4in
        \headheight 12pt \topskip .4in
        \textheight 6.9in \footskip 0pt
        \def\@oddhead{\thepage\hfil\addtocounter{page}{1}\thepage}
        \let\@evenhead\@oddhead \def\@oddfoot{} \def\@evenfoot{} }

\def\numberbysection{\@addtoreset{equation}{section}
        \def\theequation{\thesection.\arabic{equation}}}

\def\underline#1{\relax\ifmmode\@@underline#1\else
        $\@@underline{\hbox{#1}}$\relax\fi}

\def\titlepage{\@restonecolfalse\if@twocolumn\@restonecoltrue\onecolumn
     \else \newpage \fi \thispagestyle{empty}\c@page\z@
        \def\thefootnote{\fnsymbol{footnote}} }

\def\endtitlepage{\if@restonecol\twocolumn \else \newpage \fi
        \def\thefootnote{\arabic{footnote}}
        \setcounter{footnote}{0}}  %\c@footnote\z@ }

\catcode`@=12
\relax

\def\figcap{\section*{Figure Captions\markboth
        {FIGURECAPTIONS}{FIGURECAPTIONS}}\list
        {Figure \arabic{enumi}:\hfill}{\settowidth\labelwidth{Figure
999:}
        \leftmargin\labelwidth
        \advance\leftmargin\labelsep\usecounter{enumi}}}
 \relax
\def\tablecap{\section*{Table Captions\markboth
        {TABLECAPTIONS}{TABLECAPTIONS}}\list
        {Table \arabic{enumi}:\hfill}{\settowidth\labelwidth{Table
999:}
        \leftmargin\labelwidth
        \advance\leftmargin\labelsep\usecounter{enumi}}}
 \relax
\def\reflist{\section*{References\markboth
        {REFLIST}{REFLIST}}\list
        {[\arabic{enumi}]\hfill}{\settowidth\labelwidth{[999]}
        \leftmargin\labelwidth
        \advance\leftmargin\labelsep\usecounter{enumi}}}
 \relax
\makeatletter
\newcounter{pubctr}
\def\publist{\@ifnextchar[{\@publist}{\@@publist}}
\def\@publist[#1]{\list
        {[\arabic{pubctr}]\hfill}{\settowidth\labelwidth{[999]}
        \leftmargin\labelwidth
        \advance\leftmargin\labelsep
        \@nmbrlisttrue\def\@listctr{pubctr}
        \setcounter{pubctr}{#1}\addtocounter{pubctr}{-1}}}
\def\@@publist{\list
        {[\arabic{pubctr}]\hfill}{\settowidth\labelwidth{[999]}
        \leftmargin\labelwidth
        \advance\leftmargin\labelsep
        \@nmbrlisttrue\def\@listctr{pubctr}}}
 \relax
\makeatother
\newskip\humongous \humongous=0pt plus 1000pt minus 1000pt

\newif\ifdtup

\relax

\def\be{\begin{equation}}
\def\ee{\end{equation}}
\def\ba{\begin{eqnarray}}
\def\ea{\end{eqnarray}}

\def\del{\partial}

\def\r{\rho}

\def\om{\omega}
\def\Om{\Omega}
\def\l{\lambda}

\def\s{\sigma}

\def\qq{\qquad}

\def\IR{\relax{\rm I\kern-.18em R}}

\def \ov {\over}

\def\IR{\relax{\rm I\kern-.18em R}}
\def\IL{\relax{\rm I\kern-.18em L}}

\def\inv{^{\raise.15ex\hbox{${\scriptscriptstyle -}$}\kern-.05em 1}}

\begin{document}

\renewcommand{\theequation}{\thesection.\arabic{equation}}
\csname @addtoreset\endcsname{equation}{section}

\newcommand{\beq}{\begin{equation}}
\newcommand{\eeq}[1]{\label{#1}\end{equation}}
\newcommand{\ber}{\begin{eqnarray}}
\newcommand{\eer}[1]{\label{#1}\end{eqnarray}}
\newcommand{\eqn}[1]{(\ref{#1})}
\begin{titlepage}
\begin{center}

%\today
\hfill CPHT-RR014.0314

\vskip  0.8in

\boldmath
{\large \bf Gravity duals of $\mathcal{N}=2$ SCFTs and asymptotic\\ emergence of the electrostatic description}
\unboldmath

\vskip 0.6in

{\bf P. Marios Petropoulos,$^{1}$\phantom{x} Konstadinos Sfetsos$^{2}$}
\phantom{x}and\phantom{x} {\bf {Konstadinos Siampos}$^{3}$
}
\vskip 0.3in
{\em
${}^1$Centre de Physique Th\'eorique, Ecole Polytechnique, CNRS UMR 7644,\\
91128 Palaiseau Cedex, France\\
{\tt marios.petropoulos@cpht.polytechnique.fr}\\
\vskip .15in
${}^2$Department of Nuclear and Particle Physics,\\
Faculty of Physics, University of Athens,\\
Athens 15784, Greece\\
{\tt ksfetsos@phys.uoa.gr}\\
\vskip .15in
${}^3$M\'ecanique et Gravitation, Universit\'e de Mons Hainaut,
  7000 Mons, Belgique\\
{\tt konstantinos.siampos@umons.ac.be}\\
}
\vskip .6in

\textbf{Abstract}

\end{center}

We built the first eleven-dimensional supergravity solutions with $SO(2,4)\times SO(3)$ $ \times U(1)_R$ symmetry that exhibit the
asymptotic emergence of an extra $U(1)$ isometry. This  enables us to make the connection with the usual electrostatics--quiver
description.
The solution is obtained via the Toda frame of K\"ahler surfaces with vanishing scalar
curvature and $SU(2)$ action.

\end{titlepage}

\newpage
%\vskip .3in

\tableofcontents

\noindent

\vskip .4in
\noindent
%August 2002\\

%\vfill
%\eject

\def\baselinestretch{1.2}
\baselineskip 20 pt
\noindent

%%%%%%%%%%%%%%%

\setcounter{equation}{0}
\section{Introduction}
\renewcommand{\theequation}{\thesection.\arabic{equation}}

Finding explicit solutions of eleven-dimensional supergravity admitting dual $\mathcal{N}=2$ field theories is a challenging, though well-owed problem. The first example was presented in \cite{Maldacena:2000mw}, while general features and properties have been developed since in \cite{Lin:2004nb,Gaiotto:2009gz, ReidEdwards:2010qs, Donos:2010va, Aharony:2012tz}, making contact in particular with $\mathcal{N}=2$ quiver gauge theories.

Assuming a specific form for the metric and the antisymmetric fields, the problem boils down to finding solutions of the continual Toda equation, subject to appropriate boundary conditions. The solution of Toda equation can exhibit a symmetry, which translates at the level of the geometry into an extra $U(1)$ isometry. When this happens, the Toda problem is equivalent to solving a Laplace equation \cite{Ward:1990qt} and addresses the cylindrically symmetric electrostatic problem  of a perfectly conducting plane with a line charge distribution normal to it \cite{Gaiotto:2009gz}.

The electrostatic picture is useful for unravelling the quiver interpretation of the dual field theory. It is however a stringent limitation and it is desirable to understand more general situations without electrostatic analogue. A first step in that direction was taken in \cite{Petropoulos:2013vya}, where an explicit two-parameter family of solutions of the Toda equation without extra symmetry was exhibited. The idea underlying the construction  was to borrow solutions from other systems, where Toda equation governs the dynamics. Four-dimensional  gravitational configurations are among those, and in particular self-dual gravitational instantons of the Boyer--Finley
type \cite{Boyer}. Assuming that these are furthermore Bianchi IX foliations, Toda solutions are obtained by solving other integrable systems such as Darboux--Halphen \cite{Takhtajan:1992qb}, which are well understood irrespective of the symmetry, and using the mapping provided in \cite{Olivier:1991pa, Finley:2010hs}.

The analysis performed in \cite{Petropoulos:2013vya} is a real \emph{tour de force} in terms of finding
eleven-dimensional supergravity solutions. The solutions obtained in this way have no smearing and thus no extra $U(1)$ symmetry,  even asymptotically. This good feature in terms of novelty is altogether a caveat because it  does not provide any handle for the interpretation of the dual field theory.

In the present note, we propose another set of supergravity solutions, for which the absent $U(1)$ is restored in some asymptotic corner of the geometry. These are technically less involved than that in \cite{Petropoulos:2013vya}. They are based on solutions of Toda equation as they appear in another class of remarkable four-dimensional geometries, namely metrics with a symmetry, vanishing scalar curvature and K\"ahler structure. The specific metrics we consider here belong to the more general class of LeBrun metrics \cite{LeBrun}, and combine again the Bianchi IX  feature as it emerges in a class known as \emph{Pedersen--Poon K\"ahler surfaces with zero scalar curvature} \cite{Poon}.

\section{Scalar-flat four-dimensional K\"ahler spaces}

The purview of this section is to set-up the contact with Toda equation via the so-called K\"ahler-plus-symmetry LeBrun metrics \cite{LeBrun}
for the \emph{Pedersen--Poon} class \cite{Poon}.

The LeBrun geometries possess a $U(1)$ isometry, are K\"ahler  and have vanishing scalar curvature.
The presence of the $U(1)$ isometry, realised with the Killing vector $\partial_{\varphi}$, enables the metric to be set in the form
\begin{equation}
\label{LeBrun}
\text{d}s^2=\frac{1}{U}\left(\text{d}{\varphi}+A\right)^2+U  \gamma_{ij}\text{d}x^i \text{d}x^j\ ,
\end{equation}
where 
\begin{equation}
\label{LeBrungamma}
\gamma_{ij}\text{d}x^i \text{d}x^j =
\text{e}^\Psi(\text{d}x^2+\text{d}y^2)+\text{d}z^2,
\end{equation}
is the Toda frame and $U,\Psi$ being generically functions of $x,y$ and $z$, whereas $A$ is a one-form. Extra symmetries may in general appear and affect this dependence.

The  K\"ahler condition entails
\be
\label{Aform}
\text{d}A=\partial_x U \, \text{d}y\wedge \text{d}z+\partial_y U \, \text{d}z\wedge \text{d}x+\partial_z\left(U\, \text{e}^\Psi\right)\text{d}x\wedge \text{d}y\ ,
\ee
with integrability condition
\be
\label{eqU}
\left(\partial_x^2+\partial_y^2\right)U+\partial_z^2\left(U\text{e}^\Psi\right)=0\ ,
\ee
also known as linearised Toda equation.
Imposing in addition the vanishing of the scalar curvature $R$ gives the differential equation
\be
\label{eqpsi}
\left(\partial_x^2+\partial_y^2\right)\Psi+\partial_z^2 \text{e}^\Psi=0,
\ee
which is precisely the continual Toda.\footnote{ 
Notice that the left-hand side of the Toda equation
can be recast as $\text{e}^\Psi\nabla_3^2\Psi$, where $\nabla_3$ refers to the three-dimensional metric
\eqref{LeBrungamma}.}

One should stress that according to LeBrun \cite{LeBrun}, \emph{every} K\"ahler-plus-symmetry metric with vanishing $R$ is locally of the form \eqref{LeBrun} and \eqref{LeBrungamma}, with $A, U,\Psi$ satisfying \eqref{Aform}--\eqref{eqpsi},  and conversely  every metric in the class \eqref{LeBrun}--\eqref{eqpsi} is K\"ahler-plus-symmetry with vanishing $R$. The K\"ahler form reads:
\be
J=(\text{d}{\varphi}+A)\wedge \text{d}z-U\text{e}^\Psi \text{d}x\wedge \text{d}y,
\ee
and satisfies $\text{d}J=0$.
%
\begin{comment}
\begin{equation}
\text{d}J=0, \quad i_{\varphi} J=\text{d}z\,,
\end{equation}
which defines the coordinate $z$.
\end{comment}
%

Let us for completeness and later use remind that a four-dimensional K\"ahler metric has vanishing scalar curvature if and only if it is Weyl  anti-self-dual with respect to the canonical orientation induced by the
K\"ahler structure \cite{Itoh}. Due to the presence of this canonical orientation, the equivalence between self-dual and anti-self-dual metrics is broken. In practice this subtlety plays a role in a very limited number of instances,\footnote{These include the Fubini--Study metric on $\mathbb{C}\mathrm{P}_2=\nicefrac{SU(3)}{U(2)}$ and its non-compact counterpart, the
(pseudo-)Fubini--Study metric on $\widetilde{\mathbb{C}\mathrm{P}_2}=\nicefrac{SU(2,1)}{U(2)}$. The latter geometries are
K\"ahler--Einstein and Weyl self-dual -- the only known of this type with $SU(2)$ action \cite{Dancer.Einstein}.}
and discussing them here is out of our main goal.

K\"ahler metrics with vanishing scalar curvature can have more that one isometry. A class of geometries with at least three Killing vectors are Bianchi IX foliations, of the form:\footnote{Alternatively expressed as
$\text{d}s^2  =\Om_1\Om_2\Om_3 \text{d}\tau^2 + {\Om_2\Om_3\ov \Om_1}\s_1^2 + {\Om_3\Om_1\ov \Om_2}\s_2^2 + {\Om_1\Om_2\ov \Om_3}\s_3^2
$. Unlike the hyper-K\"ahler and quarternionic cases, for K\"ahler metrics with vanishing scalar curvature, the diagonal ansatz is not the most general one \cite{Dancer}.\label{dia}}
\be
\label{BIX}
\text{d}s^2 = {\Om_1 \Om_2\ov \Om_3} \om_1\om_1^*+ {\Om_3 \ov \Om_1 \Om_2} \om_2\om_2^*,
\ee
where
\be
\om_1 = \Om_3 \text{d}\tau + i \s_3\ ,\qq \om_2  =\Om_2 \s_1 + i \Om_1\s_2
\ee
with $\Om_i$ functions of $\tau$, and $\s_i$ the left $SU(2)$-invariant Maurer--Cartan one-forms obeying $\text{d}\s_1 = \s_2\wedge \s_3$ and cyclic.
When necessary, we will use the explicit parameterisation{
\begin{equation}
\sigma_1+i\sigma_2=-\mathrm{e}^{i\,\psi}\left(i\,\mathrm{d}\vartheta+\sin\vartheta\,\mathrm{d}\varphi\right)\,,\qquad
\sigma_3=\mathrm{d}\psi+\cos\vartheta\,\mathrm{d}\varphi
\end{equation}
with Euler angles $(\vartheta,\psi,\varphi) \in [0,\pi]\times [-2\pi,2\pi]\times [0,2\pi]$.}
This metric has generically $SU(2)$ symmetry, which can be enhanced to $SU(2)\times U(1)$ if two of the $\Omega$s are equal or to
$SU(2)\times SU(2)$ if they are all equal.

Imposing the K\"ahler condition and vanishing scalar curvature on \eqref{BIX} leads to the developments of Pedersen and Poon \cite{Poon} (the reader is redirected to the original reference for details).
The requirement of \eqref{BIX} being K\"ahler  leads to the system of first-order coupled differential equations:
\be
\Om_1' = \Om_2 \Om_3 - a \Om_1, \qquad \Om_2' = \Om_3 \Om_1 - a \Om_2,\qquad \Om_3' = \Om_1 \Om_2\ ,
\label{omcub}
\ee
where $a $ is a real function of $\tau$ and the prime stands for the derivative with respect to $\tau$. Demanding furthermore that  the scalar curvature vanishes, imposes $a $ be constant,
which we take here positive.
The resulting (manifestly closed) K\"ahler form is
\be
J = {i\ov 2} \left( {\Om_1 \Om_2\ov \Om_3} \om_1\wedge \om_1^* + {\Om_3 \ov \Om_1 \Om_2} \om_2\wedge \om_2^*\right)=
\Omega_1\Omega_2\,\mathrm{d}\tau\wedge\s_3+\,\Om_3\,\mathrm{d}\s_3\ .
\ee

Before scanning the solutions of Eqs. \eqref{omcub}, we would like to set up the dictionary for translating them into solutions of the Toda equation. This is possible since,
being K\"ahler with vanishing scalar curvature,  \eqref{BIX}--\eqref{omcub} can always be recast along the lines of \eqref{LeBrun}--\eqref{eqpsi} \cite{Tod.Scalar1}.
The transformation reads:
\be
\label{Tod.LeBrun}
\begin{split}
& U^{-1}=\Omega_1\Omega_2\Omega_3\sum_{i=1}^3 \left(\frac{n_i}{\Omega_i}\right)^2\,,\\
& A_i\,\mathrm{d}x^i=U\left(\left(\frac{\Omega_1\Omega_3}{\Omega_2}-\frac{\Omega_2\Omega_3}{\Omega_1}\right)\,\sin\vartheta\,\sin\psi\,\cos\psi\,\mathrm{d}\vartheta+\frac{\Omega_1\Omega_2}{\Omega_3}\,\cos\vartheta\,\mathrm{d}\psi\right) \ ,\\
&\Psi={-2a \tau}\,,\\
& x=\text{e}^{a\tau}\,n_1\,\Omega_1\,,\qquad y=\text{e}^{a \tau}\,n_2\,\Omega_2\,, \qquad z=n_3\,\Omega_3\,,
\end{split}
\ee
where $n_1=\cos\psi\sin\vartheta, n_2=\sin\psi\sin\vartheta$ and $n_3=\cos\vartheta$ are the directional cosines obeying $\displaystyle\sum_{i=1}^3n_i^2=1$. Furthermore, using the Jacobian of the transformation relating $(x,y,z)$ and $(\tau, \theta,\psi)$, as well as the Pedersen--Poon Eqs. \eqref{omcub}, one obtains the following relations:
\be
\label{dzpsi}
\partial_z\Psi=-\frac{2a n_3\,\Om_1\Om_2}{n_1^2\Om_2^2\Om_3^2+n_2^2\Om_3^2\Om_1^2+n_3^2\Om_1^2\Om_2^2}\,,
\ee
and
\be
\label{dzpsid}
\frac{\partial_z\Psi}{z}=-\frac{2a \Om_1\Om_2}{\Omega_3\left(n_1^2\Om_2^2\Om_3^2+n_2^2\Om_3^2\Om_1^2+n_3^2\Om_1^2\Om_2^2\right)}\ ,
\ee
which will prove useful later. Using \eqref{omcub}, one finally checks that $(A,U,\Psi)$ satisfy Eqs. \eqref{Aform}, \eqref{eqU} and \eqref{eqpsi}, respectively. As already advertised,
solving Eqs. \eqref{omcub} translates via \eqref{Tod.LeBrun} into solutions of the Toda equation.

In practice using the latter of \eqref{Tod.LeBrun}, we eliminate $(\vartheta,\psi)$ and we obtain the equation of an ellipsoid
\be
\label{ellipsoid}
\frac{x^2}{\text{e}^{2a \tau}\,\Omega_1^2}+\frac{y^2}{\text{e}^{2a \tau}\,\Omega_2^2}+\frac{z^2}{\Omega_3^2}=1\,,
\ee
which implicitly determines {$\tau$ (and the Toda potential, using $\Psi={-2a  \tau}$)} as a function of $(x,y,z).$

\section{Toda from Pedersen--Poon}

\subsection{Boundary conditions and general equations}
\label{geneq}

Our scope is now to analyse the system \eqref{omcub} and interpret  its solutions in the Toda frame.
Keeping in mind that these are meant to serve as building blocks for eleven-dimensional supergravity admitting $\mathcal{N}=2$ duals, one should be careful with their boundary conditions, and keep only those which satisfy
\begin{equation}
\label{bcN2}
\left. \partial_z\Psi\right\vert_{z \to 0}\sim z\to0\,,\qquad \left. \text{e}^\Psi\right\vert_{z \to 0}=\rm{finite}\neq0\,.
 \end{equation}
In the case of punctures, the $U(1)_R$ circle shrinks in a smooth manner if \cite{Gaiotto:2009gz,Aharony:2012tz}
\begin{equation}
\label{puncture}
z=z_c=2N_5\,,\qquad{ \left. \partial_z\Psi\right\vert_{z \to z_c}\to\infty\,,\qq \left. \text{e}^\Psi\right\vert_{z \to z_c}}\sim z-z_c\,,
\end{equation}
where $N_5$ is the number of M5-branes.

There are several branches of solutions to the system \eqref{omcub} under investigation. The simplest one has $a =0$, and the associated four-dimensional geometries are the Riemann self-dual (thus Ricci-flat) gravitational instantons found by Eguchi--Hanson \cite{Eguchi:1978gw, Eguchi:1979yx} and
generalised in \cite{Belinsky:1978ue}. It is known that their Toda potential is trivial, as one can readily see from \eqref{Tod.LeBrun}.
Therefore we will assume that $a \neq0$, and study separately two distinct cases, according  to their symmetries. In the first, the symmetry is enhanced and we recover the known electrostatic analogy; in the second, the symmetry remains unaltered, and we provide new solutions.

The best way to perform the analysis is to recast the system \eqref{omcub} into a single second-order differential equation. It is convenient to introduce a
new coordinate $t$ as
\begin{equation}
\label{Omega.biaxial-r}
 at  =\text{e}^{-a  \tau}\,.
\end{equation}
We learn from the first two Eqs.  \eqref{omcub} that
\begin{equation}
\label{FInew}
s\equiv\frac{1}{t^2a ^2}\left(
\Omega_1^2-\Omega_2^2
\right)
\end{equation}
is a first integral. If non-zero, its value is irrelevant because it can be reabsorbed in a redefinition of $t$; so either $s=0$ or $s=1$. This enables us to parametrise the functions $\Omega_i$ in terms of a single function $w(t)$ as follows:
\begin{equation}\label{Om12s}
\Omega_1=\frac{a  t }{2}\left(w+\frac{s}{w}\right)\, , \qquad \Omega_2=\epsilon \frac{a  t }{2}\left(w-\frac{s}{w}\right)\, , \qquad \epsilon =\pm 1\, .
\end{equation}
When $s=0$, $\Omega_1=\epsilon \Omega_2$ and the isometry of \eqref{BIX} is enhanced to $SU(2)\times U(1)$, where the last factor is generated by $\partial_\psi$; this configuration is called \emph{biaxial}.
In the instance where $s=1$, the symmetry is $SU(2)$ and the solution is called \emph{triaxial}. Hence, the Toda equation will have an electrostatic analogue for $s=0$ only. The option $\epsilon = \pm 1$ in \eqref{Om12s} deserves a comment. As one can see from  \eqref{BIX} (or its form given in footnote \ref{dia}), the four-dimensional metric is equally well-defined with positive or negative $\Omega$s -- up to an overall sign -- provided their signs do not change along $\tau$ (or $t$). The allowed range of variation for the latter is thus defined by demanding that every $\Omega_i$ keeps its sign unaltered. From the eleven-dimensional perspective, the range of allowed $t$ is mostly dictated by the limits set with \eqref{bcN2} and \eqref{puncture}.

Using the system  \eqref{omcub}, one finds the differential equation obeyed by $w$:
\begin{equation}
\label{painl}
\frac{\mathrm{d}^2w}{\mathrm{d}t^2}=\frac{1}{w}
\left(\frac{\mathrm{d}w}{\mathrm{d}t}\right)^2-\frac{1}{t}\frac{\mathrm{d}w}{\mathrm{d}t}+
\frac{w^3}{4}-\frac{s}{4w}\,,
\end{equation}
whereas $\Omega_3$ is given by
\begin{equation}
\Omega_3=-\epsilon \frac{a  t }{w}\frac{\text{d}w}{\text{d}t}\, .\label{Om3}
\end{equation}
Equation \eqref{painl} is Painlev\'e\footnote{The general Painlev\'e III equation is
$$
\frac{\mathrm{d}^2w}{\mathrm{d}t^2}=\frac{1}{w}
\left(\frac{\mathrm{d}w}{\mathrm{d}t}\right)^2-\frac{1}{t}\frac{\mathrm{d}w}{\mathrm{d}t}
+
\frac{1}{t}\left(\alpha  w^2 +\beta\right)
+
\gamma w^3+\frac{\delta}{w}\,.
$$
} III with $(\alpha,\beta,\gamma,\delta)=\left(0,0,\nicefrac{1}{4},\nicefrac{-s}{4}\right)$. It has remarkable features that will be useful in the subsequent analysis. Notice that by setting $w=\exp G$, this equation is mapped onto the central-symmetric two-dimensional Liouville  ($s=0$) or  sinh-Gordon ($s=1$) equations:
 \begin{equation}
 \frac{1}{t}\frac{\text{d}}{\text{d}t}\left(t\frac{\text{d}}{\text{d}t}G\right)=\frac{1}{4}\left(\text{e}^{2G}- s\,\text{e}^{-2G}\right).
\end{equation}

Before proceeding with the separate analysis of biaxial and triaxial solutions, a few generic remarks should be made here. From \eqref{Tod.LeBrun} and \eqref{Omega.biaxial-r} we obtain:
\begin{equation}
\text{e}^{\Psi}=(ta)^2,
\end{equation}
which vanishes at $t=0$, and is otherwise finite. Hence, \emph{punctures can only emerge at the locus $t=0$} provided $\partial_z\Psi$ diverges. We also recall from \eqref{Tod.LeBrun} that
\begin{equation}
\label{zthet}
z=\cos \vartheta \, \Omega_3(t).
\end{equation}
The latter vanishes $\forall t$ at $\vartheta=\nicefrac{\pi}{2}$, which should be interpreted as a coordinate artefact, as well as at any value $t_*$ such that $\Omega_3(t_*)=0$. Condition \eqref{bcN2} should be fulfilled at these points.

Finally, solutions to Painlev\'e III equation are algebraic or transcendental. In either case, they systematically possess poles (or branch points) at $t_a$, sometimes in infinite number inside $\mathbb{C}$. On the real axis, a \emph{bona fide}  solution $w$ will set intervals $(t_a, t_{a+1})$, which naturally restrict the range for the coordinate $t$. On the one hand, within such an interval, $w$ may have an extremum, and thus $\Omega_3$ a root (following \eqref{Om3}), while generically $\Omega_{1,2}$ remain finite and thus  $\partial_z\Psi$ vanishes (see \eqref{dzpsi} and \eqref{zthet}). According to \eqref{bcN2}, this invalidates the solution. On the other hand,  $w$ may vanish at $t_*$, making $\Omega_3$ diverge, and $\Omega_{1,2}$ vanish or diverge depending on $s$ (see \eqref{Om12s}). This behaviour is acceptable, but further  restricts the interval to $(t_a, t_*)$ or $(t_*, t_{a+1})$.

\boldmath
\subsection{Enhanced $SU(2)\times U(1)$ symmetry and electrostatics}
\unboldmath
\label{sec:biax}

Lets us consider the biaxial situation, and set for concreteness $\epsilon = 1$ in Eqs. \eqref{Om12s} and
\eqref{Om3} --  the case  $\epsilon = -1$ does not bring any physically new input.
The equation of Painlev\'e III now at hand is  algebraically integrable, with general solution
\begin{equation}
\label{w.biaxial}
w=4\kappa \frac{\nicefrac{\zeta}{a}}{(\kappa t)^{1-\nicefrac{\zeta}{a}}-(\kappa t)^{1+\nicefrac{\zeta}{a}}},
\end{equation}
where $\zeta$ and $\kappa$ are two arbitrary constants. There is always a pole or a branch point (depending on the actual value of $\nicefrac{\zeta}{a}$) at $t=\nicefrac{1}{\kappa}$. The value of $\kappa$ is otherwise irrelevant and we will set it equal to 1. Furthermore, $w$ is invariant under $\zeta \to -\zeta$, and the parameter space is therefore reduced to $\zeta>0$.
From Eqs. \eqref{Om12s} and  \eqref{Om3}, using \eqref{w.biaxial} we obtain:
\begin{equation}
\label{Omega.biaxial}
\Omega_1=\Omega_2=2\zeta\frac{t^{\nicefrac{\zeta}{a}}}{1-t^{\nicefrac{2\zeta}{a}}}
\,,\qquad \Omega_3=a-\zeta \frac{1+t^{\nicefrac{2\zeta}{a}}}
{1-t^{\nicefrac{2\zeta}{a}}}
\, .
\end{equation}
The corresponding four-dimensional K\"ahler metric with vanishing scalar curvature \eqref{BIX} is known as LeBrun metric.

In the case under consideration, there are two natural intervals for $t$: $(0,1)$ and $(1,+\infty)$. In the range $(1,+\infty)$, no $t$ makes $w$ extremal, and this interval is \emph{a priori} acceptable for any $\zeta$. For $t \in (0,1)$, however, we must impose that $\zeta\geqslant a$ to avoid vanishing $\Omega_3$ at $t_*>0$ (extremum of $w$).

We can refine this analysis by calling for the alternative electrostatic picture. Remember that the extra $U(1)$ isometry originates from the choice of a foliation \eqref{BIX} over three-spheres that are homogeneous and axially symmetric (because $\Omega_1=\Omega_2$\footnote{The same holds for $\Omega_1=-\Omega_2$.}). It also emerges in the Toda frame, where $\Psi(x,y,z)$
is effectively a function of two coordinates only:  $r=\sqrt{x^2+y^2}$  and $z$.

Let us for completeness show how this description arises in general, following
\cite{Ward:1990qt} and the analysis performed in \cite{Gaiotto:2009gz,ReidEdwards:2010qs,Petropoulos:2013vya}. The Toda potential $\Psi(r,z)$ satisfies
Eq. \eqref{eqpsi}, which simplifies:
\be
\label{eqpsi-el}
\frac{1}{r}\partial_r (r \partial_r \Psi)+\partial_z^2 \text{e}^\Psi=0\, .
\ee
In this case, we can map the Toda potential $\Psi$ to an electrostatic potential $\Phi$. This requires trading $(r,z)$ for
$(\rho,\eta)$ as
\begin{equation}
\label{Ward}
\ln r = \partial_\eta \Phi\ ,\qquad z = \rho\partial_\rho \Phi \ ,\qquad \rho=r \mathrm{e}^{\nicefrac{\Psi(r,z)}{2}}\,,
\end{equation}
which, together with \eqref{eqpsi-el}, leads for
$\Phi=\Phi(\rho,\eta)$ to the equation
\begin{equation}
\label{Laplacian}
\frac{1}{\rho}\partial_\rho(\rho\partial_\rho\Phi)+\partial^2_\eta\Phi=0\,.
\end{equation}
This is the scalar Laplacian equation in cylindrical coordinates $(\rho,\eta)$.

We can now apply the above for an axisymmetric Bianchi IX foliation. The ignorable coordinate is $\psi$, and the coordinates $(t, \vartheta)$ are ultimately replaced with $(\rho, \eta)$, via $(r,z)$.
Using \eqref{Tod.LeBrun} and \eqref{Ward}, one finds:
\begin{equation}
\label{PPaxi1}
\rho=\left\vert\Omega_1\right\vert \sin\vartheta
\,,\qquad \eta=\cos\vartheta(\Omega_3-a)\,,
\end{equation}
where $\Omega_{1,3}$ are displayed in \eqref{Omega.biaxial}. The electrostatic potential finally reads:
\begin{equation}
\label{PPaxi2}
\Phi(\rho, \eta)=\eta\ln\left(\frac{\rho}{ta}\right)+a\left(\cos\vartheta+\ln\tan\frac{\vartheta}{2}\right)\,,
\end{equation}
where $t$ and $\vartheta$ are implicit functions of $(\rho, \eta)$, obtained by inverting \eqref{PPaxi1}.

Equations \eqref{PPaxi1} and \eqref{PPaxi2} provide the electrostatic picture of Pedersen--Poon axisymmetric solution \eqref{Omega.biaxial}, describing some K\"ahler Bianchi IX foliation with zero scalar curvature. We can recast the boundary conditions for $\Psi$, Eqs. \eqref{bcN2} and \eqref{puncture}, in electrostatic language as well as in terms of the $\Omega$s, and compare with the already quoted literature \cite{Lin:2004nb, Gaiotto:2009gz, ReidEdwards:2010qs,Donos:2010va, Aharony:2012tz, Maldacena:2000mw}.

The locus $z=0$ in \eqref{bcN2} leads to $\del_\r \Phi \big |_{\eta=0} = 0$ or $\rho=0$.
This actually reflects a boundary condition: $\Phi $ being an electrostatic potential, the surface $\eta=0$ appears as an infinite conducting  plane, and
\be
\label{charge}
\l(\eta) \equiv \r\del_\r \Phi\big |_{\r =0} = z(\r=0,\eta)
\ee
as a line charge density along the $\eta$-semiaxis.\footnote{More rigorously, Eq. \eqref{Laplacian} should be
$\partial_\rho\left(\rho\partial_\rho\Phi\right)+\rho\,\partial^2_\eta\Phi= \lambda (\eta)\delta(\rho)$.} Since we know $\Phi$ (Eq.  \eqref{PPaxi2}),  we can readily find $\lambda(\eta)$ and,
using Eqs. \eqref{Tod.LeBrun}, \eqref{Omega.biaxial-r} and \eqref{PPaxi1}--\eqref{charge}, express it in terms of the original Pedersen--Poon data. This can be performed in the two distinct ranges of $t$ quoted above, potentially corresponding to two different eleven-dimensional solutions:
\begin{description}
\item[\boldmath $t\in (1, +\infty)$\unboldmath] At $\eta=0$, \emph{i.e.} on the infinite conducting plane, the range $\rho \in (0,+\infty)$ covers  $t\in (+\infty, 1)$. At large $t$, $\Omega_1=\Omega_2$ vanish as $-t^{\nicefrac{-\zeta}{a}}$ (see \eqref{Omega.biaxial}), whereas $\Omega_3$ reaches its asymptotic value $a+\zeta$. Combining all the data one finds:
\begin{equation}
\label{dis-sym}
  \lambda(\eta) =
     \begin{cases}
       \frac{a+\zeta}{\zeta}\,\eta\, ,\quad 0 \leqslant \eta\leqslant\zeta \quad (\frac{\pi}{2} \geqslant\vartheta \geqslant 0\ \& \ t\to +\infty)
       \\
  \eta+a\, ,\quad \zeta\leqslant\eta
  \quad (\vartheta =0\ \& \ +\infty > t > 1)
  \,.
     \end{cases}
\end{equation}
Regularity of the corresponding eleven-dimensional supergravity solution (originally charge conservation) also demands \cite{Gaiotto:2009gz} the reduction of slope at $\eta =\zeta$ be of 1 unit. Thus $a=\zeta$. The change of slope must furthermore occur at integer values of $\eta$, enforcing thereby $a$ be a positive integer.  In summary, the eleven-dimensional interpretation brings supplementary constraints with respect to the original Pedersen--Poon four-dimensional, K\"ahler scalar-flat space:
\begin{equation}
\label{electro.rules}
a=\zeta\in\mathbb{N}^*\,,\qquad \vartheta\in[0,\pi/2].
\end{equation}

\item[\boldmath $t\in (0,1)$\unboldmath]
In this case we are restricted to the range $\zeta\geqslant a$. On the conducting plane $\eta=0$, $t$ varies from $0$ to $1$ while $\rho$ increases from $0$ to $+\infty$. At $t=0$, $\Omega_1=\Omega_2=0$ and $\Omega_3=a-\zeta$. We now obtain for the line-charge density:
\begin{equation}
\label{dis-sym-b}
  \lambda(\eta) =
     \begin{cases}
       \frac{\zeta-a}{\zeta}\,\eta\, ,\quad 0 \leqslant \eta\leqslant\zeta \quad (\frac{\pi}{2} \leqslant\vartheta \leqslant \pi\ \& \ t=0)
       \\
  \eta-a\, ,\quad \zeta\leqslant\eta
  \quad (\vartheta =\pi\ \& \ 0 \leqslant  t < 1)
  \,.
     \end{cases}
\end{equation}
Punctures might be present in the range $0 \leqslant \eta\leqslant\zeta$, where $t=0$ and $z=(a-\zeta)\cos\vartheta$. However, this configuration lacks regularity because the slope increases from the first branch to the second.  The only way out is to set $a=\zeta=0$, which trivializes the solution.
\end{description}

In conclusion, the first biaxial solution obtained using Pedersen--Poon procedure \eqref{dis-sym} is regular but resembles the $\text{AdS}_7\times S^4$
solution. Although the second one \eqref{dis-sym-b}  is degenerate, it has the virtue to suggest that moving to the triaxial configurations may leave some  freedom for accommodating regularity, while recovering the electrostatics in some corner of the space.

\boldmath
\subsection{Strict $SU(2)$ symmetry and new solutions}
\unboldmath

We now set $s=1$ in Eq. \eqref{painl}, and deal with the triaxial problem, where generically $\Omega_1\neq\Omega_2\neq\Omega_3$ are given in Eqs. \eqref{Om12s} and \eqref{Om3}; again $\epsilon=1$  for concreteness.\footnote{Notice that choosing $\epsilon=-1$ is equivalent to trading $w$ for $\nicefrac{1}{w}$, while keeping $\epsilon =1$. Painlev\'e III with $(\alpha,\beta,\gamma,\delta)=\left(0,0,\nicefrac{1}{4},\nicefrac{-1}{4}\right)$ in invariant under $w \to\nicefrac{1}{w}$, hence if $w$ is a solution, so is $\nicefrac{1}{w}$.} Painlev\'e III is no longer algebraically integrable. Its solution is a Painlev\'e III transcendent, which is, as usual, better described in terms of its movable singularities (poles or branch points), rather than in terms of initial conditions.  The interested reader can find precious information about these properties in \cite{gromak}, or in the literature on sinh-Gordon equation as e.g.
\cite{Jaworski}. The useful properties for our subsequent analysis can be summarised as follows:
\begin{itemize}
\item The solutions have an infinite number of simple poles in $\mathbb{C}$.
\item At large $t$, $\vert w\vert$ is exponentially decreasing.
\item At small $t$, the behaviour is
\begin{equation}
\label{stbeh}
w=\frac{\kappa}{t^{\zeta}}\left(1 + \mathcal{O}\left(t^2\right)
\right)\, , \quad 0\leqslant \zeta<1\, .
\end{equation}
\end{itemize}

The large-$t$ region is not so appealing for two reasons. Firstly, according to the general discussion of the end of Sec. \ref{geneq}, we do not expect any puncture in this regime. Secondly, at large $t$, $\Omega_1$ and $\Omega_2$ do not converge towards each other because $\Omega_1-\Omega_2=\nicefrac{at}{w}$ diverges exponentially. We therefore miss the potential contact with the biaxial regime. Nevertheless, solutions to Painlev\'e III equation can make sense from the eleven-dimensional perspective. Indeed, $\exp \Psi$ is regular, and when $t$ decreases from infinity, $\vert w\vert$ increases, until it hits $\vert w\vert=1$, for some $t_*$. There, either $\Omega_1$ or  $\Omega_2$ vanishes, and this sets the acceptable domain for the eleven-dimensional solution: $(t_*, +\infty)$.

The small-$t$ regime is more interesting. Indeed, $\exp \Psi=(at)^2$ vanishes at $t=0$, potential location of punctures, and $\Omega_1-\Omega_2\propto t^{1+\zeta} \left(1 + \mathcal{O}\left(t^2\right)\right)\approx 0$ in this neighborhood, \emph{restoring thereby the extra $U(1)$ symmetry}. 
More precisely, using \eqref{Om12s}, \eqref{Om3} and  \eqref{stbeh}, we obtain:
\begin{eqnarray}
\Omega_1\approx \Omega_2&=&\frac{a\kappa}{2}t^{1-\zeta} \left(1 + \mathcal{O}\left(t^2\right)\right)\, , \\
 \Omega_3&=& a \zeta + \mathcal{O}\left(t^2\right)\, .
\end{eqnarray}
We conclude that at $t=0$, $z=a\zeta \cos \vartheta$ (see \eqref{zthet}). This excludes
the limiting case $\zeta=0$, for if $\zeta=0$, $\left.z\right\vert_{t=0}=0$, and  this cannot be the location of punctures (see \eqref{bcN2}). For $0<\zeta<1$ we can check the condition \eqref{puncture}, and use it for determining the exact location of the punctures. We find from Eq. \eqref{dzpsi}:
\begin{equation}
\left.\partial_z\Psi\right\vert_{t=0}=-\frac{2\cos\vartheta}{a\zeta^2 \sin^2\vartheta},
\end{equation}
which diverges at $\vartheta=0$, whereas $\vartheta=\pi$ is disregarded due to the expectation $z_c>0$. The punctures are thus located at $(t=0, \vartheta=0)$ \emph{i.e.} at $z=z_c$, where $z_c=a\zeta>0$. 

Our conclusion is that the Painlev\'e III transcendants at hand provide a  Pedersen--Poon configuration, corresponding, via the Toda frame, to a regular
eleven-dimensional supergravity solution with
\begin{equation}
N_5=\frac{a\zeta}{2}\,,\qquad\vartheta\in[0,\nicefrac{\pi}{2}].
\end{equation}
This solution being triaxial, it has just $SO(2,4)\times SO(3)\times U(1)_R$ isometry.

As anticipated at the end of Sec. \ref{sec:biax}, although biaxial Pedersen--Poon solutions that incorporate punctures are not available, triaxial configurations do exist. Moreover, the extra $U(1)$ biaxial symmetry is restored, in these solutions, in the vicinity of the punctures, at $z=z_c$. This is the main achievement of the present letter.

\section{Conclusion and outlook}

The scope of this note was to generalise the results of \cite{Petropoulos:2013vya}, where
the first family of eleven-dimensional supergravity solutions, dual to four-dimensional
SCFTs, and with everywhere strict $SO(2,4)\times SO(3)\times U(1)_R$ isometry was constructed. The generalisation we presented here, exhibits an asymptotic emergence of the extra $U(1)$ symmetry, that if it were present everywhere, would allow for a genuine electrostatic description. This asymptotic emergence sets the bridge with previous works on electrostatics \cite{Gaiotto:2009gz,ReidEdwards:2010qs,Donos:2010va,Aharony:2012tz}, and may turn useful for unravelling the nature of the dual gauge theories of our supergravity configurations.

Our construction is based of the Toda frame for four-dimensional K\"ahler surfaces with vanishing scalar curvature, LeBrun spaces \cite{LeBrun} specialised to Bianchi IX foliations \cite{Poon}. The extra $U(1)$ isometry is realised around the punctures. Understanding the consequences of the existence of this region deserves further investigation, in particular from  the perspective of the dual gauge field theory. The latter is expected to be a non-perturbative quiver, but the arguments in favor of this interpretation are too primitive to be exposed here.

\section*{Acknowledgements}
The authors benefited from discussions and exchanges with J.P.~Derendinger, B. Grammatikos, V.~Niarchos, Ph.~Spindel, A. Ramani, A.~Tomasiello and A.~Zaffaroni. The research of P.M. Petropoulos was  supported by the LABEX P2IO, the ANR contract 05-BLAN-NT09-573739 and the ERC Advanced Grant 226371.
The research of K.\,Sfetsos is implemented
under the \textsl{ARISTEIA} action (D.654 GGET) of the \textsl{Operational
programme education and lifelong learning} and is co-funded by the
European Social Fund (ESF) and National Resources (2007-2013). The work of K.
Siampos has been supported by the \textsl{ARC -- Direction g\'en\'erale de
l'Enseignement non obligatoire et de la Recherche scientifique -- Communaut\'e fran\c{c}aise de
Belgique} (AUWB-2010-10/15-UMONS-1), and by IISN-Belgium (convention 4.4511.06). The authors
would like to thank each others home institutions for hospitality
and financial support.
%%%%%%%%%%%%%%%%%%%%%%%%%%%%%%%%%%%%%%%%%%%%%%%%

\end{document}